\documentclass[aps,superscriptaddress,11pt,twoside]{revtex4}
\usepackage{latexsym,amsmath,amsfonts,amssymb}
\usepackage{graphicx,epsfig}
\usepackage{amsthm}
\usepackage{verbatim}
\usepackage[nohug]{diagrams}
\diagramstyle[labelstyle=\scriptstyle]

\def\R{\mathbb{R}}

\def\C{\mathbb{C}}
\def\Z{\mathbb{Z}}

\def\pmx{\begin{pmatrix}}                 % ver 1.2 Nicolas
\def\emx{\end{pmatrix}}                   % ver 1.2 Nicolas
\def\N{\mathcal{N}}                 % quantum channel
\def\Nc{\mathcal{N}^c}
\def\M{\mathcal{M}}                 % quantum channel

\def\D{\mathcal{D}}
\def\C{\mathcal{C}}
\def\A{\mathcal{A}}

\def\H{\mathcal{H}}
\def\C{\mathcal{C}}

\def\S{\mathcal{S}}
\def\B{\mathcal{B}}

\def\U{\mathcal{U}}

\def\p{\mathcal{P}}
\def\cR{\mathcal{R}}

\def\N{\mathcal{N}}                      % quantum channel
\def\Nc{\mathcal{N}^c}              % complementary channel
\def\D{\mathcal{D}}                      % degrading map
\def\Da{\mathcal{D}^a}              % antidegrading map
\def\Nn{\mathcal{N}^{\otimes n}}    % n-fold use of channel
\def\Cl{\mathcal{C}l}               % cloning channel
\def\id{\mathbb{I}}                 % identity                            %v0.5
\def\n{\hat{n}}                     % unit Bloch vector n
\def\pauli{\vec{\sigma}}               % pauli matrices

\newcommand{\Li}[2]{{\mathrm{Li}}\left(#1,#2\right)}

\def\ra{\rightarrow}

\newtheorem{theorem}{Theorem}
\newtheorem{lemma}{Lemma}
\newtheorem*{rem}{Remark}

\newtheorem*{conj}{Conjecture}

\def\Tr{\operatorname{Tr}}

\def\bra#1{\mathinner{\langle{#1}|}}
\def\ket#1{\mathinner{|{#1}\rangle}}

\newcommand{\kb}[2]{| #1\rangle\!\langle #2 |}

\def\aj{\alpha_j}
\def\ajj{\alpha_{j+1}}
\newcommand{\smfrac}[2]{\mbox{$\frac{#1}{#2}$}}
\def\Id{\operatorname{id}}
\newcommand{\proj}[1]{| #1\rangle\!\langle #1 |}

\begin{document}

\title{Conjugate Degradability and the Quantum Capacity of Cloning Channels}
%\author{Kamil Br\'adler, Nicolas Dutil, Patrick Hayden, Abubakr Muhammad}

\author{Kamil Br\'adler}
 \email{kbradler@cs.mcgill.ca}
 \affiliation{
    School of Computer Science,
    McGill University,
    Montreal, Quebec, H3A 2A7, Canada
    }

\author{Nicolas Dutil}
 \email{ndutil@cs.mcgill.ca}
 \affiliation{
    School of Computer Science,
    McGill University,
    Montreal, Quebec, H3A 2A7, Canada
    }

\author{Patrick Hayden}
 \email{patrick@cs.mcgill.ca}
 \affiliation{
    School of Computer Science,
    McGill University,
    Montreal, Quebec, H3A 2A7, Canada
    }

\author{Abubakr Muhammad}
 \email{abubakr@lums.edu.pk}
 \affiliation{
    School of Computer Science,
    McGill University,
    Montreal, Quebec, H3A 2A7, Canada
    }
 \affiliation{
    School of Science \& Engineering,
    Lahore University of Management Sciences,
    Opposite Sector U, DHA, Lahore - 54792, Pakistan
    }

\date{17 September 2009}

\maketitle

\section*{Abstract}
A quantum channel is conjugate degradable if the channel's
environment can be simulated up to complex conjugation using the
channel's output. For all such channels, the quantum capacity
can be evaluated using a single-letter formula. In this article we
introduce conjugate degradability and establish a number of its
basic properties. We then use it to calculate the quantum capacity
of $N$ to $N+1$ and $1$ to $M$ universal quantum cloning machines as
well as the quantum capacity of a channel that arises naturally when
data is being transmitted to an accelerating receiver. All the
channels considered turn out to have strictly positive quantum
capacity, meaning they could be used as part of a communication
system to send quantum states reliably.

\section{Introduction} \label{sec:intro}

The quantum capacity of a noisy quantum channel measures the
channel's ability to transmit quantum coherence. The capacity is
defined by optimizing over all possible quantum error correcting
codes in the limit of many uses of the channel and with the
provision that decoding errors ultimately become negligible. Because
of the ubiquity of noise in quantum systems, error correction will
necessarily play a central role in future attempts to build devices
exploiting quantum information, which makes finding a general
expression for the quantum capacity, and the codes associated with
it, one of the most important problems in quantum information
theory.

To date, the best general lower bound on the capacity is given by
optimizing a quantity known as the (one-shot) coherent
information~\cite{SchumacherN96,Lloyd96,Shor02,Devetak05}.
Unfortunately, it is also known that for some channels this
prescription fails to yield capacity~\cite{ShorS96,DiVincenzoSS98}.
For at least one class of channels, however, known as the
\emph{degradable} channels, the prescription \emph{does} yield the
quantum capacity and, therefore, the optimal
codes~\cite{DevetakS05}. Degradable channels were introduced by
Devetak and Shor, building on ideas from classical information
theory~\cite{Bergmans73}. Roughly speaking, they are channels for
which the output of the channel always contains at least as much
information as the channel's environment, in the sense that the
channel environment can be simulated using the channel output. Many
familiar channels are in fact degradable, such as generalized
dephasing channels, erasure channels with erasure probability at
most $1/2$ and Kraus diagonal channels~\cite{CubittRS08}. If the
role of the channel output and the environment are reversed in the
definition, the channel is called \emph{antidegradable}. By a simple
no-cloning argument, such channels always have zero quantum
capacity. A host of familiar channels are in this class, including
all entanglement-breaking channels. Indeed, nearly every channel for
which the quantum capacity is known is either degradable or
antidegradable.

In this article, we expand the list of quantum channels for which
the quantum capacity can be calculated by introducing the property
of \emph{conjugate degradability}. Conjugate degradability is
defined much like degradability except that the channel output can
be used to simulate the environment only up to complex conjugation
or, equivalently, transposition. Many desirable properties of
degradability continue to hold despite this modification, including,
for example, the fact that the optimized coherent information is
equal to the quantum capacity.

Conjugate degradability first appeared implicitly in
\cite{BradlerHP08}, which analyzed private data transmission in the
presence of a uniformly accelerated eavesdropper. When the data is
encoded using a bosonic, dual-rail qubit, the effective channel to
the eavesdropper is a conjugate degradable channel with
infinite-dimensional output. Not all examples are so exotic,
however. Optimal $N \rightarrow N+1$ and $1\rightarrow M$ cloning
machines are also conjugate degradable, as we will demonstrate
before using the property to evaluate their capacities.
Understanding coherence preservation in these channels is
particularly interesting given the well-known connection between
detecting separability and the existence of symmetric
extensions~\cite{DohertyPS02}; cloning channels attempt to construct
symmetric extensions for arbitrary states. We find a strictly
positive general formula for the quantum capacities of $N \ra N+1$
and $1 \ra M$ universal quantum cloning machine (UQCM), establishing
that all such machines are useful for transmitting quantum
information reliably.

{\bf Structure of the paper:} Section \ref{sec:conj_basic}
introduces the notion of conjugate degradability and demonstrates
that there is a single-letter formula for the quantum capacity of
conjugate degradable channels. Section \ref{sec:examples} uses
conjugate degradability to evaluate the quantum capacity of Unruh
channels as well as the universal quantum cloning machines. Next,
section \ref{sec:gen_prop} establishes a number of structural
properties of conjugate degradable channels following the template
for degradable channels established by Cubitt \emph{et
al.}~\cite{CubittRS08}. Finally, in an appendix, we present some
supporting evidence in favor of the conjecture that all $N \ra M$
universal quantum cloning machines are conjugate degradable.

\section{The quantum capacity of conjugate degradable channels}\label{sec:conj_basic}

\subsection{Notation}

Before we proceed, let us recall some relevant concepts from quantum
information theory and fix some notation. Let $\B(\H)$ represent the
space of bounded linear operators on the Hilbert space $\H$. A
quantum channel is a completely positive, trace-preserving (CPTP)
map between two such spaces of operators. We will typically call the
input space $A$ and the output space $B$ so that the quantum channel
is a map $\N : \B(A) \rightarrow \B(B)$. The identity channel will be denoted by $\id$.

Every quantum channel has a \emph{Stinespring dilation}, a
representation of the channel in which the action of the channel on
an input density matrix $\varrho$ is given by $\N(\varrho) = \Tr_E(U
\varrho U^\dagger)$, where $U : A \rightarrow B \otimes E$ satisfies
$U^\dagger U = I$. $E$ labels an auxiliary space usually called the
environment because it models the effect of noise on the channel.
There is also a complementary channel $\Nc: \B(A) \ra \B(E)$, given
by tracing over the output space instead of the environment:
$\Nc(\varrho) = \Tr_B(U \varrho U^\dagger)$.

Throughout, we will use subscripts to label subsystems, so that
$\varrho_A = \Tr_B \varrho_{AB}$, for example. The density operator
of a pure state $\ket{\psi}$ will sometimes be denoted by $\psi$.
$H(\tau) = -\Tr (\tau \log \tau)$ is the von Neumann entropy of a
density matrix $\tau$. Sometimes it is more convenient (or
important) to emphasize the particular system on which the density
matrices act. In that case, we will write $H(X)_\tau = H(\tau_X)$ to
denote the  von Neumann entropy of the state $\tau$ restricted to
the space $X$. The conditional entropy $H(X|Y)_\tau$ is defined to
be $H(XY)_\tau - H(Y)_\tau$.

The \emph{coherent information} of a quantum channel for a given
input density matrix is given by $I_c(\N, \varrho) = H(\N(\varrho))
- H(\Nc(\varrho)).$ Equivalently, given the Stinespring dilation $U$
for $\N$ and setting $\tau = U\varrho U^\dagger$, $I_c(\N,\varrho) =
H(B)_\tau - H(E)_\tau$. As is evident from the formula, the coherent
information measures the reduction via leakage into the environment
of the information transmitted about $\varrho$ . The quantum
capacity of a channel $\N$ is given by the limit
\begin{equation}\label{eq:QC1}
    Q(\N) = \lim_{n\ra \infty} \frac{1}{n} Q^{(1)}(\Nn),
\end{equation}
where $Q^{(1)}(\N)$ is the maximum output coherent information of a
channel, given by
\[
    Q^{(1)}(\N) = \max_\varrho I_c(\N, \varrho).
\]

\subsection{Degradable quantum channels}

We can now introduce the \emph{degradability} of quantum channels. A
channel $\N$ is called \emph{degradable} if there exists a CPTP map,
$\D:\B(B)\ra \B(E)$ such that $\D \circ \N = \Nc$. Similarly, a
channel is \emph{antidegradable} if its complementary channel is
degradable, or equivalently, if there exists a CPTP map $\Da$ such
that $\Da \circ \Nc = \N$. This has been shown in the accompanying
diagram:
\begin{diagram}[textflow]
A &\rTo^{\N} &   B \\
  &\rdTo_{\Nc} &   \dDotsto^{\D} \uDotsto_{\Da} \\
  & &  E
\end{diagram}
As mentioned above, degradable channels are interesting because
their quantum capacities are easily evaluated. The reason for this
ease is the observation in \cite{DevetakS05} that for degradable
channels, the optimized coherent information is additive. In other
words, $Q^{(1)}(\Nn) = n Q^{(1)}(\N)$. This means that the quantum
capacity in (\ref{eq:QC1}) can be computed using the
\emph{single-letter} formula $Q(\N)= Q^{(1)}(\N) = \max_\varrho
I_c(\N, \varrho)$, as compared to the general case in which
essentially nothing is known about the rate of convergence to the
limit.

\subsection{Conjugate degradability}\label{subsec:conj_basic}

In \cite{BradlerHP08}, the authors investigated a channel with
infinite dimensional output that was degradable up to complex
conjugation of the output space. In other words, the channel wasn't
obviously degradable but it nearly was. In this paper we generalize
that example to the class of {conjugate degradable} channels,
showing not only that there exist some important finite dimensional
examples in this class but that the conjugate degradable channels
share many of the useful properties of degradable channels with
respect to their structure and capacities.

To understand conjugate degradability, it is helpful to consider the
following diagram:

\begin{diagram}[textflow]
  &            &            &  B               &            &  \\
  &            & \ruTo^{\N}  &  \dDotsto_{\D} &  \rdTo^{\C} &        \\
  & A           &            &  E'            &             & B' \\
  &            & \rdTo_{\Nc} &  \uTo_{\C}     &  \ruDotsto_{\Da}&          \\
  &            &              &  E            &              &
\end{diagram}
As above, $A,B,E$ play the role of the input, output and the
environment. Let $\C: \B(E)\ra \B(E')$ denote entry-wise complex
conjugation with respect to a fixed basis of $E \cong E'$. A quantum
channel is called \emph{conjugate degradable} if there exists a CPTP map
$\D: \B(B) \ra \B(E')$ such that the diagram commutes. In other
words,
\begin{equation}\label{eq:conjDeg1}
    \D \circ \N = \C \circ \Nc.
\end{equation}
In a similar fashion, $\N$ is \emph{conjugate antidegradability} if there exists a CPTP conjugate degrading map
$\Da$ for the complementary channel giving
\begin{equation}\label{eq:conjaDeg1}
    \Da \circ \N^c = \C \circ \N.
\end{equation}
It is immediate from the definition that $\N$ is conjugate degradable if and only if $\Nc$ is conjugate antidegradable.

Let us now consider what happens during the computation of the
quantum capacity for conjugate degradable channels. The map $\D:
\B(B) \ra \B(E')$ can also be represented by its Stinespring
dilation with the addition of another auxiliary space, say $F$, so
that we have a corresponding isometry $V: B \ra E'F$ in addition to
the isometry $U:A \ra BE$ for $\N$. Thus $H(B)_\tau = H(E'F)_\omega$
for the states $\tau = U \varrho U^\dagger$ and $\omega = V \tau
V^\dagger$. Now consider the coherent information of the channel,
$I_c(\N, \varrho) = H(B)_\tau-H(E)_\tau = H(E'F)_\omega-H(E')_\omega
= H(F|E')_\omega$, using the fact that complex conjugation does not
change the entropy. For two uses of the channel $\N^{ \otimes 2}:
\B(A_1A_2) \ra \B(B_1B_2)$, repeated application of the strong
subadditivity of von Neumann entropy yields
\begin{eqnarray}
    H(F_1F_2|E_1'E_2')_\omega  &\leq& H(F_1|E_1'E_2')_\omega + H(F_2|E_1'E_2')_\omega \nonumber \\
                        &\leq& H(F_1|E_1')_\omega + H(F_2|E_2')_\omega \nonumber
\end{eqnarray}
where $\omega = (V \otimes V)\tau(V^\dagger \otimes V^\dagger)$ and
$\varrho = (U \otimes U)\varrho_{A_1 A_2}(U^\dagger \otimes U^\dagger)$
for any $\varrho_{A_1 A_2}$. This shows that $I_c(\N^{ \otimes 2},
\varrho_{A_1A_2}) \leq I_c(\N, \varrho_{A_1}) + I_c(\N,\varrho_{A_2})$, which
in turn implies that $Q^{(1)}(\N^{\otimes 2}) \leq 2 Q^{(1)}(\N)$.
Since the superadditivity is an immediate consequence of the
definition and we can easily deduce by induction that $Q^{(1)}(\Nn)
\leq n Q^{(1)}(\N)$, we clearly have, just as in the case of
degradable channels, a single-letter expression for the capacity of
a conjugate degradable channel,
\begin{equation}\label{eq:QC2}
  Q(\N) = Q^{(1)}(\N) = \max_\varrho I_c(\N, \varrho).
\end{equation}

Next, consider the class of conjugate antidegradable channels.
Suppose that it is possible to send quantum states through such a
channel in the sense that there is a decoding CPTP map $\cR$ such
that $(\cR \circ \N)(\phi)$ has high fidelity with $\ket{\phi}$ for
all $\ket{\phi}$ in a subspace of dimension at least 2. By the
definition of conjugate antidegradability, the output
$\Nc(\kb{\phi}{\phi})$ of the complementary channel can be
transformed to a high-fidelity copy of the conjugated state
$\ket{\bar\phi}$ using the antidegrading map $\Da$. Combining these
two operations, from $A$ to $B$ and from $A$ to the conjugated $B'$
via the complementary channel, approximately implements the map
$\ket{\phi} \mapsto \ket{\phi} \ket{\bar\phi}$. But this
transformation is nonlinear and is therefore not possible. It
follows that the transmission of the quantum states with high
fidelity over the antidegradable quantum channel is not possible.
Likewise, the quantum capacity of all conjugate antidegradable
channels is identically zero.

%%%% begin Ver 1.2 / Sept 25, 2008
Referring back to the previous diagram, if $\N$ is conjugate
degradable, then for any state $\varrho_{AR}$,
\[
    (\C \otimes \rm{id}) (\Nc \otimes \rm{id}) \varrho_{AR} = (\D \circ \N \otimes \rm{id}) \varrho_{AR} \geq 0.
\]
%Without loss of generality, take $\varrho_{AR}$ to be the maximally
%entangled state or the singlet given by
%$\frac{1}{\sqrt{n}}\sum_{i=1}^n \ket{i}\ket{i}$.
Thus, the state $\sigma_{ER} = (\Nc \otimes \rm{id}) \varrho_{AR}$ has
a positive partial transpose. (Complex conjugation and transpose act identically on Hermitian matrices.) From \cite{HHH96}, there are only two
possibilities:
\begin{enumerate}
  \item $\sigma_{ER}$ is a convex combination of product states (also known as a separable state), which is equivalent to saying that $\Nc$ is entanglement-breaking. Since entanglement-breaking channels are a subclass of antidegradable channels, this implies that $\N$ is degradable.
  \item $\sigma_{ER}$ is a bound entangled state, which is equivalent to saying that $\Nc$ is entanglement-binding.
\end{enumerate}

Moreover, bound entanglement does not exist in quantum systems of total dimension less than or
equal to six~\cite{HHH96}. As a result,  conjugate
antidegradable channels with input
and output dimensions satisfying this constraint
are also entanglement-breaking. As a special
case, suppose $\N$ is a conjugate degradable qubit channel (input and output dimension equal to two). We will prove in Theorem \ref{thm:qubitoutput} that the Choi rank of conjugate degradable qubit
channels is less than or equal to two. This means that $\Nc$ is also
a qubit channel, which implies that the product of $\Nc$'s input and
output space dimensions is less than six. But since $\Nc$ is also
conjugate antidegradable, $\Nc$ has to be entanglement-breaking.

%%%% end Ver 1.2 / Sept 25, 2008

\section{Examples} \label{sec:examples}

\subsection{Example 1: $1\rightarrow 2$ universal quantum cloning machine} \label{subsec:12cloner}

While cloning quantum states is impossible, approximate cloning is permitted provided the quality of the clones is sufficiently poor. A $1 \ra 2$ universal quantum cloner is a CPTP map $\Cl : \B(A) \ra \B(B_1B_2)$, where $A, B_1$ and $B_2$ are all qubits such that
\begin{itemize}
\item $\Tr_{B_1} \Cl = \Tr_{B_2} \Cl$: the outputs on $B_1$ and $B_2$ are identical.
\item The Uhlmann fidelity~\cite{Uhlmann76,Jozsa94} $F( \psi, \Tr_{B_1} \Cl (\psi) )$ is independent of the input state $\psi$.
\end{itemize}
The optimal cloning machine maximizes the fidelity $F( \psi, \Tr_{B_1} \Cl (\psi) )$.
Bu{\v z}ek and Hillery~\cite{BuzekH98} constructed such a machine (shown to be optimal in~\cite{BrussDEFMS98}), which we will see provides a natural example of a conjugate degradable channel.
Later we will study the more complicated case of $N \ra M$ cloning.

Let $\ket{\psi} = \alpha \ket{0}+ \beta \ket{1}$.
%Using the notational convention of~\cite{GisinM97},
The Stinespring dilation of $\Cl$ acts as follows:
\begin{equation}\label{12UQCM output}
       U :  \ket{\psi}_A  \mapsto \sqrt{\frac{2}{3}}\ket{\psi}_{B_1}\ket{\psi}_{B_2} \ket{\bar\psi}_E + \sqrt{\frac{1}{6}}\ket{\psi^\perp}_{B_1}\ket{\psi}_{B_2} \ket{\bar\psi^\perp}_E + \sqrt{\frac{1}{6}}\ket{\psi}_{B_1}\ket{\psi^\perp}_{B_2} \ket{\bar\psi^\perp}_E,
\end{equation}
where $\ket{\bar\psi} = \bar\alpha \ket{0}+ \bar\beta \ket{1},\ket{\psi^\perp} = -\bar\beta \ket{0} + \bar\alpha \ket{1}$ and $\ket{\bar\psi^\perp} = -i\sigma_Y\ket{\psi}=-\beta \ket{0} + \alpha \ket{1}$. Taking the partial trace over either $B_2 E$ or $B_1 E$ yields the density operator of the individual clones:
\begin{equation}\label{12UQCM_outB}
    \varrho_{B_j} = \frac{5}{6}\kb{\psi}{\psi} + \frac{1}{6}\kb{\psi^\perp}{\psi^\perp}.
\end{equation}
The output of the complementary channel to the environment is given by
\begin{equation}\label{12UQCM_outE}
    \varrho_E = \frac{1}{3}\kb{\bar\psi^\perp}{\bar\psi^\perp} + \frac{2}{3}\kb{\bar\psi}{\bar\psi}.
\end{equation}
Another way to visualize this transformation is that the input density matrix $\kb{\psi}{\psi} = \frac{1}{2}(\id + \n \cdot \pauli )$ gets mapped to
\begin{equation}\label{12UQCM tracedoutput}
    \varrho_{B_j} = \frac{1}{2}\Big(\id + \frac{2}{3} \n \cdot \pauli\Big), \quad
    \varrho_E = \frac{1}{2}\Big(\id - \frac{1}{3} \big(-n_x\sigma_x+n_y\sigma_y-n_z\sigma_z\big)\Big),
\end{equation}
where $\n \in \R^3$ is the unit Bloch vector and $\pauli$ is a vector
of the Pauli matrices $\sigma_x, \sigma_y, \sigma_z$. From this, it
is straightforward to build a conjugate degrading map from the output ${B_1}$ to the
environment $E$ by first depolarizing to shrink the Bloch vector by a factor of
$2$, followed by  complex conjugation. Although this does not imply degradability because of the needed
antiunitary operation of conjugation, it does prove that this
channel is conjugate degradable. (Below we will show that $\Cl$ is actually also degradable.)

We now compute the capacity of this channel. Setting $\tau = U \psi U^\dagger$, Eq.~(\ref{eq:QC2})
states that the quantum capacity is equal to $\max_\psi  [H({B_1}{B_2})_\tau - H(E)_\tau] $. Again writing $V : B \ra E'F$ for the Stinespring dilation of the conjugate degrading map $\D$ and $\omega = V \tau V^\dagger$, we also saw that the capacity could be rewritten as
\[
    \max_\psi [H(FE')_\omega - H(E')_\omega] = \max_\psi H(F|E')_\omega.
\]
For a fixed input $\psi$, let $\omega(\psi) = VU\psi U^\dagger
V^\dagger$ denote the dependence of $\omega$ on this input. Due to
the unitary covariance of the channel and the concavity of the
conditional entropy,
\begin{equation}\label{eq:max_coh}
    H(F|E')_{\omega({\psi)}} = \int_{\U(A)} H(F|E')_{\omega({W\psi W^\dagger})} dW \leq H(F|E')_{{\omega({ \int W\psi W^\dagger} dW)}} = H(F|E')_{\omega(\id/2)},
\end{equation}
where $\id/2$ is the maximally mixed state at the input. Thus the
coherent information is maximized when the input is maximally mixed.
Substituting into the formula reveals the quantum capacity of the $1\to2$ cloner to be
\begin{equation} \label{eq:12_capacity}
 Q(\Cl) = H(B_1B_2)_{\Pi^+/3} - H(E)_{\id/2} = \log 3 - 1,
\end{equation}
where $\Pi^+$ is the projector onto the symmetric subspace of $B_1B_2$.

As indicated earlier, $\Cl$ is actually also degradable, which gives
an alternative method for justifying the use of the single-letter
capacity formula. Demonstrating degradability is somewhat more
delicate than conjugate degradability, however. For the purposes of
illustration, we sketch the argument here. The universal not
($U{NOT}$) operation is the antiunitary transformation that negates
a qubit's Bloch vector. From Eq.~(\ref{12UQCM tracedoutput}), the
necessary degrading map will have the form $\D = \Tr_{B_2} \circ
U{NOT} \circ \S \circ \p^+$ where $\S$ is a depolarizing map
shrinking the Bloch vector by a factor of $2$ and $\p^+(\varrho) =
\Pi^+ \varrho \Pi^+$. ($\D$ is uniquely determined to be
$(\Cl^c)\circ \Cl^{-1}$ on the range of $\Cl$, to which the map is
effectively restricted by the inclusion of $\p^+$.) The
Jamio{\l}kowski representation~\cite{Jamiolk72,choi75} can then be
used to test the complete positivity of $\D$. Letting $\ket{\Phi} =
\smfrac{1}{2} (\ket{00}_{B_1 B_1'} + \ket{11}_{B_1 B_1'}) \otimes
(\ket{00}_{B_2 B_2'} + \ket{11}_{B_2 B_2'})$, a tedious but
straightforward calculation reveals that $(\D \otimes \Id_{B_1'
B_2'})(\Phi)$ is positive semidefinite. In fact, the map is only
just positive semidefinite: if the Bloch vector had been shrunk by a
factor less than 2 then $\D$ would not be completely positive.

\subsection{Example 2: Unruh channel}

Quantum field theory predicts that the vacuum state as defined by an inertial observer will be a thermal state from the point of view of a uniformly accelerated observer, an observation known as the Unruh effect~\cite{Unruh76}. Ref.~\cite{BradlerHP08} examined the transformation induced on a dual rail bosonic qubit channel when the receiver is uniformly accelerated, finding that the transformation can be represented by a channel
\begin{equation}
\N(\proj{\psi}) = (1-z)^3 \bigoplus_{k=0}^\infty z^k \varrho_k
\end{equation}
where $z \in (0,1)$, with increasing $z$ corresponding to increasing acceleration. If $\proj{\psi} = \smfrac{1}{2}(\id + \hat{n} \cdot \vec{\sigma})$ then
\begin{equation}
\varrho_k = \frac{(k+1)\id^{(k+2)}}{2} + \hat{n} \cdot \vec{J}^{(k+2)}
\end{equation}
where $\vec{J}^{(k+2)} = ({J}_x^{(k+2)},{J}_y^{(k+2)},{J}^{(k+2)}_z)$ are generators of
the $(k+2)$-dimensional representation of $SU(2)$.

\begin{figure}[t]
\begin{center}
\resizebox{12cm}{8cm}{\includegraphics{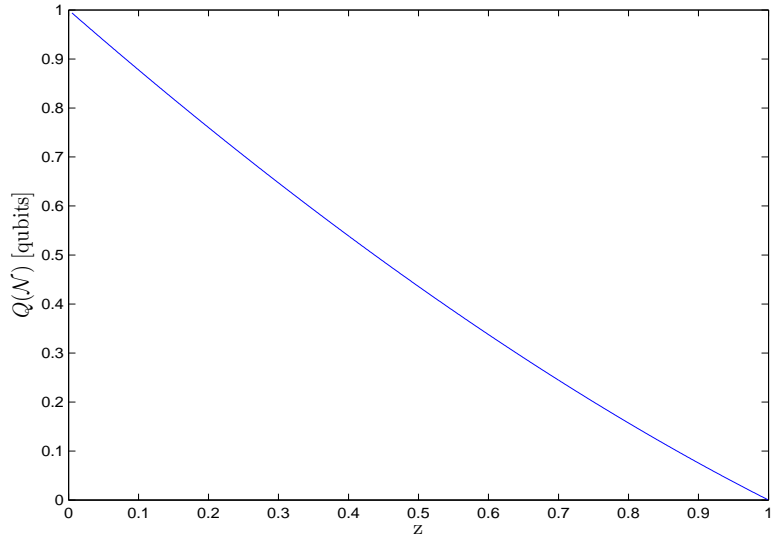}}
\caption{Quantum capacity of the Unruh channel. Acceleration is monotonically increasing with $z$. The quantum capacity decreases from $1$ for no acceleration to $0$ for infinite acceleration, exhibiting a slight convexity as a function of $z$.}
\end{center}\label{fig:quantcap}
\end{figure}

The channel was shown to be conjugate degradable and unitarily covariant
 in~\cite{BradlerHP08}. Therefore, just as for the $1 \ra 2$ UQCM, the quantum capacity will be equal to $I_c(\N,\id/2)$. Setting $\tau_B = \N(\id/2)$ and $\tau_E = \Nc(\id/2)$  we find that
\begin{equation}\label{unruhchannel_output}
%    \tau_B=\sum_{k=0}^\infty\sum_{l=0}^{k+1}T_kS_{k}\kb{H_{kl}}{H_{kl}},\hspace{1cm}
 %   \tau_E=\sum_{k=0}^\infty\sum_{l=0}^{k}T_k\tilde S_{k}\kb{\tilde H_{kl}}{\tilde H_{kl}},
    \tau_B=\bigoplus_{k=0}^\infty T_k S_{k} \id^{(k+2)} \quad \mbox{and} \quad
     \tau_E=\bigoplus_{k=0}^\infty T_k\tilde S_{k} \id^{(k+1)},
\end{equation}
where $T_k=(1-z)^3z^k$, $S_k=\smfrac{1}{2}(k+1)$ and $\tilde S_k=\smfrac{1}{2}(k+2)$.
The diagonal form of Eq.~(\ref{unruhchannel_output}) is suitable for evaluating the von Neumann entropies in $I_c(\N,\id/2)$:
\begin{eqnarray*}\label{entropy_sigma_unruh}
 Q(\mathcal{N})
  &=& H(B)_\tau -H(E)_\tau \\
  &=& -\sum_{k=0}^\infty\sum_{l=0}^{k+1}T_kS_k(\log T_k+\log S_k)+\sum_{k=0}^\infty\sum_{l=0}^{k}T_k\tilde S_k(\log T_k+\log\tilde S_k)\\
   &=& {1\over2}\sum_{k=0}^\infty T_k(k+1)(k+2)(\log\tilde S_k-\log S_k) \\
   &=& {(1-z)^3\over2}\sum_{k=0}^\infty z^k(k+1)(k+2)(\log(k+2)-\log(k+1))\\
   &=& {(1-z)^3\over2}\sum_{k=0}^\infty{{\rm d}^2\over{\rm
     d}z^2}\left[z^{k+2}\log{(k+2)}-zz^{k+1}\log{(k+1)}\right] \\
    &=&{(1-z)^3\over2}{\partial^2\over\partial z^2}{\partial\over\partial s}
    \left[(z-1)\Li{s}{z}\right]_{s=0},
\end{eqnarray*}
where $\Li{s}{z}$ is the polylogarithm function. The final expression gives a compact formula for the quantum capacity as a function of $z$ or, equivalently, acceleration.  Figure \ref{fig:quantcap} plots this function.

%In the following section, we will see that the above example of conjugate degradability can be studied %under the wider context of quantum cloning.

\subsection{Example 3: $N \ra M$ universal quantum cloning machines}\label{sec:quant_clon}

An $N \ra M$ universal quantum cloning machine $\Cl_{N,M}$ is a generalization of the $1 \ra 2$ UQCM $\Cl$ described earlier. The input system $A$ consists of $N$ identical qubits. That is, $A$ is the subspace of an $N$-qubit system $A_1 A_2 \cdots A_N$ invariant under the natural action of the symmetric group $S_N$. The output system is the $M$-qubit space $B = B_1 B_2 \cdots B_M$, where $M \geq N$, and the cloner satisfies:
\begin{itemize}
\item $\Tr_{B_j} \Cl = \Tr_{B_1} \Cl$ for all $j$.
\item The Uhlmann fidelity~\cite{Uhlmann76,Jozsa94} $F( \psi, \Tr_{B_1} \Cl (\psi) )$ is independent of the input state $\psi$.
\end{itemize}
Gisin and Massar constructed optimal $N \ra M$ cloning machines~\cite{GisinM97} with the full optimality proof appearing in~\cite{BrussEM98}.

\begin{theorem}\label{thm:clonerisconjdeg}
    All $N\to N+1$ and $1\to M$ universal quantum cloning machines for qubits are conjugate degradable.
\end{theorem}
\begin{proof}{\em (i) $N\to N+1$ case.} Tracing over any $N$ of $N+1$ output qubits we find the following density matrix~\cite{werner}
\begin{equation}\label{cloneoutput}
     \varrho_{B_i}=\eta_{N,N+1}\kb{\psi}{\psi}+{(1-\eta_{N,N+1})\over2}\id,
\end{equation}
where $\eta_{N,N+1}={N\over N+1}{N+3\over N+2}$. On the other hand, the environment is a single qubit system and of the following form
\begin{equation}\label{UNOToutput}
    \varrho_{E}=\upsilon_N\kb{\bar\psi}{\bar\psi}+{(1-\upsilon_N)\over2}\id,
\end{equation}
where $\upsilon_N=N/(N+2)$. If we undo the complex conjugation and compare $\upsilon_N$ with the channel output
we see that for all $N$ (fixed)  $\eta_{N,N+1}>\upsilon_N$ holds. This means that there always exists a CP map transforming $\varrho_{B_i}$ into $\C(\varrho_{E})\equiv\varrho_{E'}$ and thus the optimal $N\to N+1$ cloning machine is conjugate degradable.

{\em (ii) $1\to M$ case.}  We know that every UQCM is by definition
a unitarily covariant map. If, for a given input state, we find a
conjugate degrading map which is also unitarily covariant, we have a
map working for all input states and the existence of a conjugate
degrading has been proven. Let us fix the input qubit as
$\ket{\psi}=\ket{0}$. Write $\ket{N-j:0,j:1}$ for a normalized
completely symmetric state (CSS) of $N$ qubits with exactly $N-j$ of
them occupying state $\ket{0}$. Letting $j$ range from 0 to $N$
yields an orthonormal basis for the completely symmetric subspace,
which has dimension $N+1$. The action of the Stinespring dilation of
$\Cl_{1,M}$ reads~\cite{GisinM97}
\begin{multline}\label{GisMas_cloning_machine}
    \ket{1:0}_A\xrightarrow{U_{1,M}}\sum_{j=0}^{M-1}\sqrt{\alpha_j(1,M)}\ket{M-j:0,j:1}_B\ket{M-j-1:0,j:1}_E\\
    \equiv\sum_{j=0}^{M-1}\sqrt{\alpha_j(1,M)}\ket{s_j}_B\ket{r_j}_E,
\end{multline}
where $\sqrt{\alpha_j(1,M)}={M-j\over\triangle_M}$ (see Eq.~(\ref{alfa}) in the appendix) and $\triangle_M=M(M+1)/2$ is a triangle number. Therefore,
\begin{equation}\label{schmidtform1toM}
    \varrho_B={1\over\triangle_M}\bigoplus_{j=0}^{M-1}(M-j)\kb{s_j}{s_j}\quad \mbox{and}\quad \varrho_E={1\over\triangle_M}\bigoplus_{j=0}^{M-1}(M-j)\kb{r_j}{r_j}.
\end{equation}
%Note that for the $B$ subsystem we completed the completely symmetric basis ($j=M$) even though the corresponding alpha coefficient is zero.
Since consecutive eigenvalues of these matrices differ by a constant
increment, we can write both of them as linear combinations of the
identity and a diagonal generator of the $su(2)$ algebra for the
appropriate dimension.
% Eqs.~(\ref{schmidtform1toM}) have a neat
%form. Since the two consecutive eigenvalues differ by a constant
%increment we can write both of them as a sum of a maximally mixed
%(normalized) density matrix and a diagonal generator of the $su(2)$
%algebra for the given dimension.
First, we will need this
decomposition for the $E$ subsystem so we write
\begin{equation}\label{decomposition_complchannel}
    \varrho_E={\id^{\left(M\right)}\over\triangle_M}+a_MJ_z^{\left(M\right)}.
\end{equation}
A swift calculation gives us $a_M=1/\triangle_M$. Next, we introduce a new system $\tilde B$ which is given by tracing over any qubit of the $B$ system. The state is supported by a CSS basis so by this procedure we get again a state  supported by a lower-dimensional CSS basis. Moreover, the diagonality of the state from the $B$ system is preserved. Hence, following the general case given by Eq.~(\ref{tracingaqubit}) from the appendix we get
\begin{equation}\label{cloning channel output after tracing over a qubit}
    \varrho_{\tilde B}={1\over\triangle_M}\bigoplus_{j=0}^{M-1}{M^2+M-1-j(2+M)\over M}\kb{r_j}{r_j}.
\end{equation}
Note that  $\varrho_{\tilde B}$ lives in the same Hilbert space as $\varrho_E$ and the diagonal coefficients of  $\varrho_{\tilde B}$ are given by Eq.~(\ref{beta}) from the appendix.

It is intuitively  clear that when going from $B$ to $\tilde B$ we transformed a $J_z$ generator into another (lower-dimensional) $J_z$ generator. If the intuition is not convincing we can see that the difference between two consecutive eigenvalues of Eq.~(\ref{cloning channel output after tracing over a qubit}) is equal to ${1\over\triangle_M}{2+M\over M}$ and indeed independent on $j$. We may therefore again write
\begin{equation}\label{decompositioncloning channel output after tracing over a qubit}
    \varrho_{\tilde B}={\id^{\left(M\right)}\over\triangle_M}+b_MJ_z^{\left(M\right)}.
\end{equation}
From~(\ref{cloning channel output after tracing over a qubit}) and
(\ref{decompositioncloning channel output after tracing over a
qubit}) we find $b_M={1\over\triangle_M}{M^2+M-2\over M(M-1)}$ and
hence $a_M/b_M<1,\forall M>1$. Now what if the input state in
Eq.~(\ref{GisMas_cloning_machine}) is a general qubit? The output
space are not spanned by $\ket{r_j},\ket{s_j}$ anymore. The complete
output state of the transformation is explicitly dependent on the
input state as can be seen from Eq.~(\ref{cloning_machine}) for
$N=1$. Now we have to pay a special attention to the $E$ subsystem.
First of all, we introduce a complex conjugated system $E'$ which is
spanned by $\{\ket{R_j}\}_{j=0}^{M-N}$. Recall that the
transformation $\varrho_{E}\mapsto\varrho_{E'}$ is not a unitary
operation. But the subsystem $E'$ is exactly the system where the
output of a conjugate degrading map should appear (cf.
Sec~\ref{subsec:conj_basic}) and it is also the output space of the
depolarizing map defined above due to its unitary covariance. We may
conclude that we found an input-independent conjugate degrading map
$ \varrho_{\tilde B}\mapsto\varrho_{E^\prime}$.
 \end{proof}
\begin{conj}\label{THEconjecture}
    Universal quantum cloning machines  for qubits are conjugate degradable.
\end{conj}

\smallskip

%\begin{rem}
    The appendix contains some arguments supporting the conjecture.
%\end{rem}

\smallskip

To calculate the quantum capacity of $\Cl_{N,M}$ in these cases it therefore suffices to maximize the single-letter coherent information of formula (\ref{eq:QC2}). In the case of the $1 \ra 2$ UQCM, a convexity argument showed that optimum was achieved for the maximally mixed input. A similar argument applies here. The only difference is that instead of transforming according to the defining representation of $SU(2)$, the input $A$ now transforms according to a different irreducible representation $r$ of $SU(2)$. Let $\ket{\psi} \in A$ and $V : B \ra E'F$ be the Stinespring dilation of the conjugate degrading map. As before, let $\omega(\psi) = VU\psi U^\dagger V^\dagger$. For a qubit state $\ket{\phi}$, unitary covariance gives
\begin{eqnarray*}
  H(F|E')_{\omega({\phi^{\otimes N})}}
  &=& \int_{SU(2)} H(F|E')_{\omega({r(W)\phi^{\otimes N} r(W)^\dagger})} dW \\
  &\leq& H(F|E')_{{\omega({ \int r(W)\phi^{\otimes N} r(W)^\dagger} dW)}} \\
  &=& H(F|E')_{\omega(\Pi^N/(N+1))},
\end{eqnarray*}
where $\Pi^N$ is the projector onto the symmetric subspace of $N$ qubits. The unitary covariance of the channel and the fact that the output representation is also irreducible implies that $\Cl_{N,M}(\Pi^N/(N+1))$ is itself maximally mixed and therefore equal to $\Pi^M/(M+1)$. It therefore suffices to determine the output $\xi = \Nc(\Pi^N/(N+1))$ of the complementary channel.
%Using Vandermonde's identity
%\begin{equation}\label{vandermonde}
 %   \sum_{k=0}^K\binom{N}{k}\binom{M-N}{K-k}=\binom{M}{K}
%\end{equation}
%we conclude that the $\sigma$ corresponding to a maximally mixed input state, which we know from above is diagonal, has constant diagonal (unnormalized) elements
%\begin{equation}\label{diagelements}
 %  \sigma_{ii}=\sum_{k=0}^K\tilde\alpha^2_{ki}={N+1\over M+1}
%\end{equation}
%so $\sigma$ is a maximally mixed state.

The Stinespring dilation of the Gisin-Massar cloning machine acts as follows~\cite{FanMW01}:
\begin{equation} \label{eq:nm_stinespring}
    U_{NM} : \ket{N-k:0 ,k:1}_A \mapsto
    \sum_{j=0}^{M-N}\sqrt{\alpha_{kj}}\ket{M-k-j:0,k+j:1}_B\ket{P_j}_E,
\end{equation}
where both $\ket{M-k-j:0,k+j:1}_B$ and $\ket{P_j}_E$ form orthogonal sets and
\begin{equation}\label{cloner_coeffs}
    \sqrt{\alpha_{kj}}=\sqrt{\frac {(M-N)!(N+1)!}{k!(N-k)!(M+1)!}}
    \sqrt{\frac {(k+j)!(M-k-j)!}{j!(M-N-j)!}}.
\end{equation}
Therefore, an input maximally mixed state transforms as
\begin{multline}\label{transformed_inputmaxmixed}
    \bigoplus_{k=0}^N\kb{N-k:0,k:1}{N-k:0,k:1}_A\mapsto\\
    \bigoplus_{k=0}^N\sum_{j=0}^{M-N}\alpha_{kj}\kb{M-k-j:0,k+j:1}{M-k-j:0,k+j:1}_B\otimes\kb{P_j}{P_j}_E.
\end{multline}
Hence tracing over the $B$ subsystem (for every $k$) results in a diagonal matrix. Evaluating the partial trace from Eq.~(\ref{eq:nm_stinespring}) reveals that the $j-$th diagonal element of the output density matrix is of the form $\xi_{jj}=\sum_{k=0}^N\alpha^2_{kj}$. Using the identity
\begin{equation}\label{ancillaalpha}
    \alpha_{kj}={N+1\over M+1}{\binom{N}{k}\binom{M-N}{j}\over\binom{M}{k+j}}
    ={N+1\over M+1}{1\over\binom{M}{N}}{\binom{k+j}{k}\binom{M-k-j}{N-k}}
\end{equation}
gives a relatively compact expression for the diagonal entries of $\xi$:
\begin{equation}\label{ancdiagcoeffs1}
    \xi_{jj}={N+1\over M+1}{1\over\binom{M}{N}}\sum_{k=0}^N\binom{k+j}{k}\binom{M-k-j}{N-k}.
\end{equation}
The sum is reminiscent of Vandermonde's identity
\begin{equation*}\label{vandermonde}
    \sum_{k=0}^K\binom{N}{k}\binom{M-N}{K-k}=\binom{M}{K}
\end{equation*}
but more complicated because of the presence of $k$ in the `numerator'.  It can nonetheless be evaluated using generating functions.
Recall the power series~\cite[p.~336]{concretemath}
\begin{equation}\label{powerser}
    {1\over(1-z)^{1+j}}=\sum_{k=0}^\infty\binom{k+j}{j}z^k=\sum_{k=0}^\infty\binom{k+j}{k}z^k.
\end{equation}
Letting $l=M-N-j$ we get
\begin{equation}\label{ancdiagcoeffs2}
    \xi_{jj}\propto\sum_{k=0}^N\binom{k+j}{k}\binom{N-k+l}{N-k}.
\end{equation}
But then we have
\begin{equation}\label{genfunproduct}
    {1\over(1-z)^{1+j}}{1\over(1-z)^{1+\l}}={1\over(1-z)^{2-N-M}}
\end{equation}
which implies via the power series expansion that $\xi_{jj}$ is
independent of $j$ so the diagonal matrix is a multiple of identity.
Examining the coefficients in more detail yields
$$
\sum_{k=0}^N\binom{k+j}{k}\binom{M-k-j}{N-k}=\binom{1+M}{N}
$$
and thus
\begin{equation}\label{ancdiagcoeffs3}
    \xi_{jj}={N+1\over M-N+1}.
\end{equation}

Having demonstrated that both the output and environment density
operators are both maximally mixed when the input is, we get our
final result:
\begin{equation}
Q^{(1)}(\Cl_{N,M}) = \log (M+1) - \log(M-N+1) = \log
\frac{M+1}{M-N+1}.
\end{equation}
This formula generalizes the one found for the $1 \ra 2$ UQCM in
Eq.~(\ref{eq:12_capacity}) and, by virtue of their conjugate
degradability, gives the exact quantum capacity for $N \ra N+1$ and
$1 \ra M$ cloners. We see that the quantum capacity is strictly
positive for all $M \geq N$: all the optimal quantum cloning
machines are capable of transmitting quantum information reliably.
Of course, as $M/N$ increases, the maximized coherent information
decreases, going to zero as $M/N$ goes to infinity. This formula
complements the calculation of the classical capacity for all $1 \ra
M$ cloning machines presented in~\cite{Bradler09}.

\section{General properties of conjugate degradable channels}
\label{sec:gen_prop}

A recent paper by Cubitt \emph{et al.} provided a detailed investigation of the properties of degradable
channels~\cite{CubittRS08}. In this section, we reproduce a number of their results, demonstrating that
conjugate degrable channels have many of the same properties. Most of the proofs are very similar to those in
the original paper with only minor modifications required to accommodate the complex conjugation.

Throughout the section, we will abbreviate $\dim(X)$ to $d_X$.
%We will also assume that channels are implemented with a minimal-sized environment $E$ so that $d_E$ is the Choi rank of the channel.

\begin{theorem}[Compare to Theorem 1 of \cite{CubittRS08}]
\label{thm:purestate} Let $\N:\B(A) \ra \B(B)$ be a channel mapping every pure state to pure state. Then, either
%of the following possibilities must hold:
 \begin{description}
   \item[(i)]
   $d_A \leq d_B$ and $\N(\varrho) = U \varrho U^\dag$, with partial isometry $U$ such that $U^{\dag}U=I_{d_A}$,  is always
   conjugate degradable with Choi rank $d_E = 1$, or
   \item[(ii)]
   $\N(\varrho)=(\Tr \varrho)\kb{\phi}{\phi}$ for all $\varrho$. In this case, the channel $\N$ is conjugate antidegradable and maps every state to
   a single fixed pure state.
 \end{description}
\end{theorem}
\begin{proof}
The channel $\N$ can be expressed in its Stinespring dilation form by $\N(\varrho) = \Tr_E U \varrho U^{\dag}$, with $U: A \ra BE$ satisfying $U^{\dag}U = I$. Let $\{\ket{\alpha_k}\}$ be an orthonormal basis of $A$. Since the channel maps
every pure state to a pure state, we have $U \ket{\alpha_k}  = \ket{\beta_k} \otimes
\ket{\gamma_k}$ for some states $\ket{\beta_k}$ and $\ket{\gamma_k}$. Isometries preserve inner
products, and so we must have $\langle \beta_j| \beta_k \rangle \langle \gamma_j| \gamma_k \rangle =
\delta_{jk}$. For any $j \neq 1$, define the state $\phi_1 \equiv \frac{1}{2}(\ket{\alpha_1} +
\ket{\alpha_j})(\bra{\alpha_1} + \bra{\alpha_j})$. If we send $\phi_1$ through the channel $\N$, we get an
output state of the form
\begin{equation}\label{eq:pureOutput}
\N(\phi_1) = \frac{1}{2}(\kb{\beta_1}{\beta_1} + \langle \gamma_1| \gamma_j \rangle \kb{\beta_1}{\beta_j} +
\langle \gamma_j| \gamma_1 \rangle \kb{\beta_j} {\beta_1} + \kb{\beta_j}{ \beta_j}) \\
\end{equation}
If $\ket{\gamma_1}$ and $\ket{\gamma_j}$ are orthogonal, the output
state is $\N(\phi_1) = \frac{1}{2}\kb{\beta_1}{\beta_1} +
\frac{1}{2}\kb{\beta_j}{\beta_j}$, which is pure if and only if
$\ket{\beta_j}=\ket{\beta_1}$. Therefore, we must have
$\ket{\beta_j}=\ket{\beta_1}$ whenever $\langle \gamma_1| \gamma_j
\rangle = 0$.

Suppose the latter does not hold. Then we have $\langle
\beta_1|\beta_j \rangle = 0$, and the output state is pure if and
only if $|\langle \gamma_1| \gamma_j \rangle | = 1$, or,
equivalently, $\ket{\gamma_j} = e^{i\theta} \ket{\gamma_1}$. Thus,
when no states $\gamma_j$ are orthogonal to $\gamma_1$, the channel
$\N$ is of the form (i) (i.e $d_A \leq d_B$ and $d_E =1$). The
complementary channel $\N^{c}$ is given by $\N^{c}(\varrho)=(\Tr
\varrho) \proj{\gamma_1}$. Since $\Tr(U\varrho U^{\dag}) = \Tr
\varrho$ for all $\varrho \in A$, the channel $\N$ is conjugate
degradable with conjugate degrading map $\D(\sigma) = (\Tr
\sigma)\proj{\gamma_1}^{T}$.

Suppose now instead that $d_E \neq 1$. From the previous paragraph,
we may assume without loss of generality that $\langle \gamma_1|
\gamma_2 \rangle = 0$. We prepare a state $\phi_2 =
\frac{1}{2}(\ket{\alpha_2} + \ket{\alpha_j})(\bra{\alpha_2} +
\bra{\alpha_j})$, and send it through the channel $\N$. To obtain a
pure output, we must have $\ket{\gamma_j} = e^{i\theta}
\ket{\gamma_2}$ for any $j$ such that $\langle
\gamma_1|\gamma_j\rangle \neq 0$ (see Eq.~(\ref{eq:pureOutput})).
But $\ket{\gamma_j}$ cannot be proportional to two orthogonal
vectors, and so we must have $\langle \gamma_1| \gamma_j \rangle =
0$ for all $j$.  The only possible CPTP map is then $\N(\varrho) =
(\Tr \varrho)\proj{\beta_1}$. The complementary channel $\N^{c}$ is
the identity channel. The channel $\N$ is conjugate antidegradable
with conjugate degrading map $\D(\sigma) = (\Tr \sigma)
\proj{\beta_1}^{T}$.
\end{proof}

\begin{lemma}[Compare to Lemma 2 of \cite{CubittRS08}]
\label{thm:smallenvironment} Let $\N: \B(A) \ra \B(B)$ be conjugate degradable, and for a pure state
$\ket{\psi_j}$ define $B_j = \mathrm{range} ~ \N(\kb{\psi_j}{\psi_j})$ and $E_j = \mathrm{range}
~\N^{C}(\kb{\psi_j}{\psi_j})$. Then, the spaces $B_j$ and $E_j$ have equal dimensions: $d_{B_j} = d_{E_j}$.
Furthermore, if the vectors $\ket{\psi_1}, \ket{\psi_2},\ldots,\ket{\psi_{m}}$ satisfy $\mathrm{span} ~
\bigcup_j B_j=B$, we also have $\mathrm{span} ~ \bigcup_j E_j = E$ provided $d_E$ is the Choi rank of the channel.
\end{lemma}
\begin{proof}
%The proof follows similar lines of reasoning found in \cite{CubittRS08}.
Let $U: A \ra BE$ be the partial isometry in the Stinespring dilation of
the channel $\N$. The first part of the lemma can be seen to hold by looking at the Schmidt decomposition of
$U\ket{\psi_j}$:
 \begin{equation}
    U\ket{\psi_j} = \sum_{k=1}^{r_j} \mu_{jk} \ket{\phi_k^j} \otimes \ket{\omega_k^j}, \\
 \end{equation}
where for each $j$ $\{\ket{\phi_k^j}\}$ and $\{\ket{\omega_k^j}\}$ are sets of orthonormal states of the $B$ and $E$ spaces
respectively. Since the states $\{\ket{\phi_k^j}\}$ (resp. $\{\ket{\omega_k^j}\}$) are eigenvectors of
$\N(\kb{\psi_j}{\psi_j})$ (resp. $\Nc(\kb{\psi_j}{\psi_j})$), we have $r_j = d_{B_j} = d_{E_j}$.

We prove the second part of the lemma by contradiction. Assume span $\bigcup_j E_j \neq E$. Then there exists a
state $\ket{\omega^\bot} \in E$ orthogonal to span$\{\ket{\omega_k^j}:j=1\ldots m,k=1\ldots r_j\}$. Since $\N$
is conjugate degradable, we have
 \begin{equation}
   \label{eq:psiD}
    \N^C(\kb{\psi_j}{\psi_j}) = \sum_{k=1}^{r_j} \mu_{jk}^2 \kb{\omega_k^j}{\omega_k^j} = \C \circ \D \circ
 \N(\kb{\psi_j}{\psi_j}) = \sum_{k=1}^{r_j} \mu_{jk}^2 (\gamma_k^{j})^T,
 \end{equation}
where $\gamma_k^j \equiv \Tr_G V \kb{\phi_k^j}{\phi_k^j} V^{\dag}$ and the operator $V: B \ra EG$ is the partial
isometry in the Stinespring dilation of the conjugate degrading channel $\D$.

From Eq.~(\ref{eq:psiD}) and the fact that $\gamma_k^j$ is positive semidefinite, we have $0 = \Tr [
(\gamma_k^j)^T \kb{\omega^{\bot}}{\omega^{\bot}}] = \Tr [\gamma_k^j \kb{\omega^{\bot}}{\omega^{\bot}}^T] $
for all $j,k$. Consequently, $(\kb{\omega^{\bot}}{\omega^{\bot}}^T \otimes I_G)
V\ket{\phi_k^j} = 0$ for all $j,k$. By hypothesis, however, any $\ket{\phi} \in B$ can be written as a superposition of the
$\ket{\phi^j_k}$. It follows that $\bra{\omega^{\bot}}\D(\kb{\phi}{\phi})^T\ket{\omega^{\bot}}=0$ for any
$\ket{\phi} \in B$. This implies that  $\bra{\omega^\bot} \N^c(\psi) \ket{\omega^\bot} = 0$ for all states $\psi$, contradicting the fact that $d_E$ is the Choi rank of $\N$. Hence, $\mathrm{span} ~ \bigcup_j E_j = E$.
\end{proof}

\begin{theorem}[Compare to Theorem 3 of \cite{CubittRS08}]
\label{thm:fullrank} Let $\N: \B(A) \ra \B(B)$ be a CPTP map which
sends at least one pure state $\ket{\psi}$ to an output state
$\varrho = \N(\kb{\psi}{\psi})$ with full rank, i.e,
$\operatorname{rank} \N(\varrho) = d_B$. If $\N$ is conjugate
degradable, we have $d_B = d_E$.
\end{theorem}

\begin{proof}
The hypothesis of Lemma \ref{thm:smallenvironment} is satisfied with
$m=1$ so that $E = \mathrm{range} ~ \N^{c}(\kb{\psi}{\psi})$. Then, using the first part of Lemma
\ref{thm:smallenvironment}, we get
\begin{equation*}
d_B = \mathrm{dim}[\mathrm{range} \phantom{"} \N(\kb{\psi}{\psi})] = d_{B_1} = d_{E_1} =
\mathrm{dim}[\mathrm{range} \phantom{"} \N^{c}(\kb{\psi}{\psi})] = d_E.
\end{equation*}
\end{proof}
\begin{theorem}[Compare to Theorem 4 of \cite{CubittRS08}]\label{thm:qubitoutput}
Let $\N:\B(A) \ra \B(B)$ be a conjugate degradable channel with qubit output. Then, the
following two properties always hold:
\begin{description} \item[(i)]
        The Choi rank of the channel $\N$ is at most two, and
     \item[(ii)]
       The dimension of the input system $\H_A$ is at most three.
  \end{description}
\end{theorem}
\begin{proof}For the first part, assume that the Choi rank is greater than two.
Then, by Theorem \ref{thm:fullrank}, every pure state must be mapped to a rank 1 output. However, Theorem
\ref{thm:purestate} combined with the fact that $\N$ is conjugate degradable then implies that the Choi rank must be
one. The second part of the proof again follow the same lines as the proof of Theorem 4
\cite{CubittRS08} but we omit it for brevity.
\end{proof}

\begin{lemma}[Compare to Lemma 17 of \cite{CubittRS08}]\label{lemm:anti}
Let $\N : \B(A) \ra \B(B)$ be a conjugate antidegradable channel and
$\Delta : \B(B) \ra \B(C)$ any CPTP map. Then the channel $\Delta
\circ \N$ is also conjugate antidegradable.
\end{lemma}

\begin{proof} The proof follows the same lines as in Lemma 17 of \cite{CubittRS08}. The
key difference is the addition of the operator $\C$ into Eq.~(A7) of
\cite{CubittRS08}. Since $\N$ is conjugate degradable, we replace
Eq. (A7) with
 \begin{equation} \label{eq:antideg}
  (\Delta \circ \C \circ \Da \circ \Tr_{D})(\Delta \circ \N)^{c}(\varrho) = \Delta \circ \N(\varrho),
 \end{equation}
where $\Da$ is a conjugate degrading map for the complementary
channel $\Nc$ and the partial trace operator $\Tr_D$ is taken over
the environment $D$ of the channel $\Delta$. Using the Kraus
representation of $\Delta$ and the fact that $\C(\varrho) =
\varrho^T$ for any density operator $\varrho$, we have
 \begin{equation}
     \begin{split}
       (\Delta \circ \C)(\varrho) &= \sum_k E_k \varrho^T E_k^{\dag} \\
                             % &= \sum_k \bigg [\overline{E_k} \varrho E_k^T \bigg ]^{T} \\
                             &= \bigg [ \sum_k (E_k^{\dag})^T \varrho E_k^{T} \bigg ] ^{T} \\
                             &= (\C \circ \Delta')(\varrho) \\
     \end{split}
 \end{equation}
where $\Delta'$ is a map with Kraus elements $\{(E_k^{\dag})^{T}\}$.
We have
\begin{equation}
  \begin{split}
\sum_k (E_k^{\dag})^T E_k^{T} &= \sum_k (E_k E_k^{\dag})^{T} = I\\
   \end{split}
\end{equation}
and, thus, $\Delta'$ is also a CPTP map. Combining this result with
Eq.~(\ref{eq:antideg}) proves that $\Delta \circ \N$ is conjugate
antidegradable.
\end{proof}
\begin{theorem}[Compare to Theorem 18 of \cite{CubittRS08}]
\label{thm:convex} The set of conjugate antidegradable channels is
convex.
\end{theorem}
\begin{proof}
Follows easily from the proof of Theorem 18 in \cite{CubittRS08} by
replacing all antidegradable channels with conjugate antidegradable
ones.
\end{proof}

The last result of this section is comparable to Theorem 5 of
\cite{CubittRS08}, which is itself distilled from an article of Wolf
and Perez-Garcia \cite{WolfP07}. Unlike the original, however, which
demonstrated that qubit channels of Choi rank 2 are necessarily
either degradable or antidegradable, it is possible for such
channels to be neither conjugate degradable nor conjugate
antidegradable.

\begin{theorem}[Compare to Theorem 5 of \cite{CubittRS08}]
\label{thm:WolfPerezlike} Let $\N$ be a qubit channel of Choi rank
two. Then
\begin{description}
   \item[(i)]
   $\N$  is entanglement breaking if and only if it is conjugate antidegradable, and
   \item[(ii)]
   $\N$ is cannot be both conjugate degradable and conjugate antidegradable.
\end{description}
\end{theorem}
\begin{proof}
First, we recall the observation of \cite{RuskaiSW02} that, up to
unitary transformations of the input and output, every Choi rank two
qubit channel has the form
\begin{equation}
\label{Qchannel}
\N(\varrho) = A_{+} \varrho A^{\dag}_{+} + A_{-} \varrho A^{\dag}_{-},\\
\end{equation}
where
\begin{equation}\label{channelN}
    \begin{split}
       A_{+} &= \left(
       \begin{array}{cc}
       \cos\alpha & 0 \\
      0 & \cos\beta \\
      \end{array}
       \right)
       \\
        \\
       A_{-} &=\left(
       \begin{array}{cc}
        0 & \sin\beta \\
         \sin\alpha & 0
        \end{array}\right).\\
    \end{split}
\end{equation}
{\bf (i)} Reformulating in terms of the Jamio\l kowski isomorphism,
the channel $\N$ is entanglement breaking if and only if for a fixed
unnormalized maximally entangled state $\ket{\Phi} = \ket{00} +
\ket{11}$,
\begin{equation}\label{entbreakcond_of_N}
    \left[(\N\otimes\Id)(\Phi)\right]^{T_1}\geq0.
\end{equation}
($T_1$ is partial transposition over the first subsystem). This
inequality is equivalent to $1-\sin^2{\alpha}-\sin^2{\beta} = 0$,
whose solutions are $\alpha=(\pi/2+k\pi)\pm\beta$, where $k\in\Z$.
We will show that an identical condition characterizes conjugate
antidegradability.

To that end, suppose the channel $\N$ is conjugate antidegradable: on the span of $\cup_{\psi} \N^c(\psi)$,
\begin{equation}\label{conjantidegchan}
    \C\circ\D^a=\N\circ{(\N^c)}^{-1},
\end{equation}
where $\D^a$ is a conjugate antidegrading map.
% (up to complex conjugation aka
%density matrix transposition in a given basis $\C$)
%and $\N^c$ is
%the complementary channel of $\N$.
Following~\cite{WolfP07}, this can be rewritten as an equation in the
corresponding Jamio\l kowski matrices:
\begin{equation}\label{conjantidegchan_Jami}
    R_{\C\circ{\D^a}}=\left(R_\N^\Gamma\left(R_{\Nc}^\Gamma\right)^{-1}\right)^\Gamma,
\end{equation}
where $R_\M=\left[(\M\otimes\Id)(\Phi)\right]$ is the Jamio\l kowski matrix of the map $\M$ and $\Gamma$ is an involution operation~\cite{WolfP07} defined as $\langle ij|R_\N^\Gamma|jk\rangle=\langle ik|R_\N|jl\rangle$. Since $\C$ is density matrix transposition, $R_{\C\circ\D^a} =
(R_{\D^a})^{T_1}$. It follows that
\begin{equation}\label{(anti)degrador_Jami}
    R_{\D^a}=\left[\left(R_\N^\Gamma\left(R_{\Nc}^\Gamma\right)^{-1}\right)^\Gamma\right]^{T_1}.
\end{equation}
A calculation then reveals that the spectrum of $R_\A$ is
\begin{equation}\label{spectraAD1}
    \lambda_{\D^a}\in\left\{1,1,\pm{\sin^2{\beta}+\sin^2{\alpha}-1\over\sin^2{\beta}-\sin^2{\alpha}}\right\}
\end{equation}
and we immediately see that the operator $\D^a$ represents a CP map if and only if the numerator of the last two eigenvalues is zero, which coincides precisely with the condition found in Eq.~(\ref{entbreakcond_of_N}) for the positivity of the partial transpose of the output of $\N$.
%Thus, the statement of the theorem follows.

{\bf (ii)} Suppose now there exists a degrading map $\D$ up to complex conjugation $\C$
\begin{equation}\label{conjdegchan}
    \C\circ\D=\Nc\circ\N^{-1}.
\end{equation}
Following the method of the previous part, we find the eigenvalues
\begin{equation}\label{spectraAD2}
    \lambda_{\D}\in\left\{1,1,\pm{\sin^2{\beta}-\sin^2{\alpha}\over\sin^2{\beta}+\sin^2{\alpha}-1}\right\}.
\end{equation}
We see that when $\D$ is a CP map the eigenvalues of $\D^a$ blow up, indicating that the map $\D^a$ cannot be defined. Likewise, when $\D^a$ is a CP map,  the eigenvalues of $\D$ blow up.
\end{proof}

\section{Discussion}

We have introduced the property of conjugate degradability and
demonstrated that conjugate degradable quantum channels enjoy a
single-letter quantum capacity formula. We then identified some
natural examples of conjugate degradable channels: an Unruh channel
as well as both $N \ra N+1$ and $1 \ra M$ universal quantum cloning
machines. For each of these examples, we then evaluated the formulas
to get simple expressions for the channels' quantum capacities. For
the Unruh channel, the quantum capacity was strictly positive for
all accelerations and for the cloning channels, it was strictly
positive for all $M \geq N$.

We also verified that many of the properties of degradable channels
discovered in~\cite{CubittRS08} also hold for conjugate degradable
channels. For example, the set of conjugate antidegradable channels
is convex and closed under composition with arbitrary quantum
channels. In fact, it is consistent with our findings that the
conjugate degradable channels form a subset of the degradable
channels. Determining whether that is the case is the most important
open problem arising from this work. Any channels that are conjugate
degradable but not degradable would necessarily have
entanglement-binding complementary channels and therefore be at
least moderately exotic.

We hope that conjugate degradability will prove to be a useful
concept regardless of the resolution of its relationship to
degradability. The conjugate degradability of the example channels
considered in this article is essentially obvious. On the other
hand, the $1 \ra M$ universal quantum cloning machines are also
known to be degradable for $M \leq 6$, but demonstrating that fact
requires a tedious calculation and it isn't clear how to generalize
the calculation to general cloning machines~\cite{Bradler09}. More
conjugate degradable channels are certainly waiting to be
discovered. We encourage readers to find them.

\section*{Acknowledgments}

The authors thank Prakash Panangaden for conversations on the
properties of Unruh channels. This work was supported financially by
the Canada Research Chairs program, CIFAR, FQRNT, MITACS, NSERC, ONR
(\#N000140811249), QuantumWorks and the Sloan Foundation.

\section*{Appendix}

This appendix provides some motivation for the conjecture following Theorem~\ref{thm:clonerisconjdeg}. It also contains some details of the calculations used in our study of cloning channels.

The action of the Stinespring dilation for the universal qubit cloning machine ($\Cl$) on $N$ identical input qubits $\ket{N:\psi}$ can be written as~\cite{GisinM97}
\begin{multline}\label{cloning_machine}
    \ket{N:\psi}_A\xrightarrow{U_{N,M}}\sum_{j=0}^{M-N}\sqrt{\alpha_j(N,M)}\ket{M-j:\psi,j:\psi^\perp}_B\ket{M-j-1:\bar\psi,j:\bar\psi^\perp}_E\\
    \equiv\sum_{j=0}^{M-N}\sqrt{\alpha_j(N,M)}\ket{S_j}_B\ket{\bar R_j}_E,
\end{multline}
where
\begin{equation}\label{alfa}
    \sqrt{\alpha_j(N,M)}=\sqrt{N+1\over M+1}\sqrt{(M-N)!(M-j)!\over(M-N-j)!M!}.
\end{equation}
We write $\ket{N-j:\psi,j:\psi^\perp}$ for the normalized completely symmetric state (CSS) of $N$ qubits with exactly $N-j$ of them occupying state $\psi$.

First, we can transform the complementary channel output by complex
conjugation $E\mapsto E'$ simply by removing the bar from $\ket{\bar
R_j}_E$. That is, we replace $\ket{\bar{R}_j} =
\ket{M-j-1:\bar\psi,j:\bar\psi^\perp}_E$ by $\ket{R_j} =
\ket{M-j-1:\psi,j:\psi^\perp}_E$.
%This results in complex
%conjugation of density matrices representing the states living in
%the $E$ subsystem, which is not a unitary operation.
Inspecting the density matrices of the $B$ and $E'$ subsystems then
reveals a kind of Schmidt form (heretofore dropping the explicit
dependence of the diagonal coefficients on $N$ and $M$):
\begin{equation}\label{schmidtform}
    \varrho_B=\bigoplus_{j=0}^{M-N}\aj\kb{S_j}{S_j},\qquad \varrho_{E'}=\bigoplus_{j=0}^{M-N}\aj\kb{R_j}{R_j}.
\end{equation}
Now trace over a qubit in the $B$ system, for comparison with the state on $E'$:
\begin{equation}\label{tracingaqubit}
    \varrho_B\to\varrho_{\tilde B}=\bigoplus_{j=0}^{M-N}\beta_j\kb{R_j}{R_j}.
\end{equation}
Some dull but straightforward algebra yields
\begin{equation}\label{beta}
    \beta_j=\aj{M-j\over M}+\ajj{1+j\over M}
\end{equation}
if we set $\alpha_{M-N+1}=0$.

We now show that, for all $N$ to $M$ UQCM, $\varrho_{\tilde B}$
majorizes $\varrho_{E'}$ in the sense that
$\vec\beta\succ\vec\alpha$. (Let $\vec\beta^\downarrow$ be the
vector with the same coefficients as $\vec\beta$ but arranged in
nonincreasing order and likewise for $\vec\alpha^\downarrow$. Then
$\vec\beta \succ \vec\alpha$ if and only if for all $k$,
$\sum_{j=1}^k \beta_j^\downarrow \geq \sum_{j=1}^k
\alpha_j^\downarrow$.)  This majorization relation implies that for
every $\varrho_{\tilde B}$ there exists a unital CPTP map
transforming it into $\varrho_{E'}$~\cite{AlbertiU82}. What we
cannot conclude from this analysis, however, is that there is a
single unital CPTP map transforming every $\varrho_{\tilde B}$ into
$\varrho_{E'}$ as was the case with the depolarizing channel for the
cloners from Theorem~\ref{thm:clonerisconjdeg}. Such a channel, if
it exists, would be the necessary conjugate degrading map.

%This is consistent with the existence of a conjugate degrading map
%taking $\varrho_{\tilde{B}}$ to $\varrho_{E'}$ for all $\psi$ that,
%for example, acts by a mixture of unitaries. It does not, however,
%prove the existence of such a map.

\begin{lemma}\label{lem:alphabetaordered}
    The coefficients $\alpha_j$ and $\beta_j$ are in decreasing order, i.e $\alpha_j>\alpha_{j+1}$ and $\beta_j>\beta_{j+1}$.
\end{lemma}
\begin{proof}
    For $\alpha_j$ it can be seen directly from Eq.~(\ref{alfa}) that
    \begin{equation}\label{alfaalfa}
    {\aj\over\ajj}={M-j\over M-N-j}.
    \end{equation}
    For $\beta_j$  we first strip Eq.~(\ref{beta}) of all factors independent of $j$ and find
    $$
    \beta_j\propto{(M-N-j)!\over(M-N-j)!}\big((M-j)^2+(M-N-j)(1+j)\big).
    $$
    Then
    $$
    {\beta_j\over\beta_{j+1}}\propto{M-j-1\over M-j-N}{(M-j)^2+(M-N-j)(1+j)\over(M-j-1)^2+(M-N-j-1)j}
    $$
    and so ${\beta_j\over\beta_{j+1}}>1$ if ${M-j-1\over M-N-j}\geq1$ which is always true.
\end{proof}
\begin{rem}
    This result is physically sensible: states similar to the input state occur with higher probability. These are the states in which fewer qubits are rotated with respect to the input state.
\end{rem}
\begin{lemma} $\vec\beta \succ \vec\alpha$
\end{lemma}
\begin{proof}
Since the coefficients of $\vec\alpha$ and $\vec\beta$ are already ordered, it suffices to calculate:
\begin{eqnarray*}
\sum_{j=0}^k \beta_j
    &=& \sum_{j=0}^k \left[ \alpha_j \frac{M-j}{M} + \alpha_{j+1} \frac{1+j}{M} \right] \\
    &=& \alpha_0 + \sum_{j=1}^k \alpha_j \left[ \frac{M-j}{M} + \frac{j}{M} \right] + \alpha_{k+1} \frac{k+1}{M} \\
    &=& \sum_{j=0}^k \alpha_j + \alpha_{k+1} \frac{k+1}{M}.
\end{eqnarray*}
Since $\alpha_{k=1} \geq 0$, we get $\sum_{j=0}^k \beta_j \geq \sum_{j=0}^k \alpha_j$, as required.
\end{proof}

\bibliographystyle{plain}
\bibliography{conjugate}

\end{document}